# On Relationship of Koopman Eigenvalues and Frequencies in Dynamic Mode Decomposition

A.K. Alekseev

*Moscow Institute of Physics and Technology*

*Moscow, Russia*

*Abstract*

The frequency estimation from the Koopman eigenvalues (phase angles) obtained via Dynamic Mode Decomposition (DMD) is addressed. Since the calculations of the frequencies from the phase angles are nonunique, the modifications of DMD for uniqueness restoration are considered. The nonlinear oscillating mode of supersonic jet, impinging the flat plate, is used as a toy problem.

## *1. Introduction*

The Dynamic Mode Decomposition (DMD) [1,2] enables the estimation of small number of dynamic modes that describe the evolution of the total fluid state. The set of flow snapshots (usually separated by fixed time interval $\Delta t$) is used as the input data for DMD. The flow dynamics may be presented as $u(t) = \sum a_i v_i e^{(\sigma_i + i\omega_i)t}$. Here, $v_i \in C$ are the right eigenvectors of operator $A$ determining the flow evolution (linear propagator) usually denoted as dynamic (Koopman) modes, the eigenvalues $\lambda_i = \sigma_i + i\omega_i$ are designated as (Koopman) eigenvalues and coefficients $a_i \in C$ are Koopman eigenfunctions.

From other viewpoint [3], DMD provides the feasibility to approximate the evolution operator (propagator) as a product of the diagonal and two rectangular matrices $A_r = \Omega_R \Lambda \Omega_L$. These matrices are produced by the collection of right eigenvectors $\Omega_R$, left eigenvectors $\Omega_L$ and eigenvalues $\Lambda$ of the operator $A$. The special case of the Koopman operator (linear, observables coincides with dynamical variables) we mark herein as the Schmid operator.

Recovering the operator $A$ from the data set $X = (x_1, x_2, x_3 ...)$ and corresponding responses $Y = (y_1, y_2, y_3 ...)$ related by the propagator $Y = AX$ may be performed using the pseudoinverse of data matrix $X$ ($X^+ = (X^* X)^{-1} X^*$) in a form $A = YX^+$ [4]. The expression $A = YX^+$ is used in a standard way in neural networks [5]. The tremendous dimension of the matrix $A$ is the natural limitation of this approach. The special structure of the data (snapshots separated by fixed time interval) enables DMD to overcome this difficulty and distinguish DMD from neural networks. Additionally, the DMD presentation $A_r = \Omega_R \Lambda \Omega_L$ qualitatively deviates from neural networks by the presence of the matrix $\Lambda$, which contains phase angles, describing an temporal evolution.

The Koopman eigenvalues may be recast as $\lambda_j = |\lambda_j| e^{\pm i\alpha_j} = |\lambda_j|(\cos\alpha_j \pm i \sin\alpha_j) = |\lambda_j| e^{\pm i\omega_j \Delta t}$, $|\lambda_j| = e^{\sigma_j \Delta t}$, where $\alpha_j$ is a phase angle and the values $\omega_j = \alpha_j / \Delta t$ are interpreted [1,2] as the frequencies. This interpretation is nonunique, so an analysis of Koopman eigenvalues from the viewpoint of algorithms for the unique and stable estimation of frequencies is the main subject of the paper.

The impingement of 2D unsteady supersonic jet on a flat plate is used as a toy problem in the present paper to illustrate Dynamic Mode Decomposition and the existing problems for the frequency estimation.

## *2. Dynamic mode decomposition*

Consider the main features of DMD in accordance with the seminal papers [1,2]. Let's consider a set of $N$ snapshots $Sn_1^N = (u_1 ... u_N)$, which are discrete approximations of the flowfields at consequent times separated by the interval $\Delta t$. The snapshot $u_i$ is considered here as a vector of the dimension $M$. The linear operator $A(\Delta t)$ (an unknown matrix of dimension $M \times M$) is assumed to exist and provide the transformation $u_{i+1} = A u_i$. For a long enough snapshot set, the eigenvalues and right eigenvectors $A v_j = \lambda_j v_j$ may be calculated and the snapshots may be presented as

$$u_k = \sum_j a_j v_j \lambda^{k-1} = \sum_j a_j v_j e^{(\sigma_j + i\omega_j)(k-1)\Delta t}, \ (k = 1, N). \tag{1}$$

The main computational problem from this viewpoint is the estimation of Koopman modes, eigenvalues, and amplitudes from the known set of snapshots $Sn_1^N$.

Let's consider DMD algorithm in accordance with [1,4]. The shifted snapshots' sets $X = Sn_1^{N-1} = (u_1 ... u_{N-1}) = (u_1 ... A^{N-2} u_1)$, $Y = Sn_2^N = (u_2 ... u_N) = A Sn_1^{N-1}$ are used, so

$$Y = A \cdot X. \tag{2}$$

Another presentation (via the companion matrix):

$$Y = X \cdot C. \tag{3}$$

The companion matrix may be considered as the transformed propagator

$$C = X^{+}AX. \quad (4)$$

Since the snapshot matrix may be not invertible, the Moore-Penrose pseudoinverse matrix is used.

The matrix $C \in R^{N\times N}$ is not symmetric one, so the right eigenvectors do not form an orthogonal basis. The set of left eigenvectors (biorthogonal to right) is necessary for a complete description.

The companion matrix spectrum enables an estimation of the Schmid operator spectrum. From

$$C = X^{+}\Omega_R \Lambda \Omega_L X \quad (5)$$

one may obtain $\Omega_R^C = X^{+}\Omega_R$, $\Omega_L^C = \Omega_L X$ and $\Omega_R = X \cdot \Omega_R^C$, $\Omega_L = \Omega_L^C X^{+}$.

The practical implementation of DMD [1] is based on SVD [6], so $X = U\Sigma V^{*}$, $X^{+} = V\Sigma^{+}U^{*}$ and $\Sigma^{+} = diag(\sigma_1^{-1},...,\sigma_r^{-1},0,...0)$.

### *3. The reduced form of the Schmid operator*

If eigenvalues, right and left eigenvectors are available, the construction of the Schmid operator in the following reduced form is feasible:

$$A_r = \Omega_R \Lambda \Omega_L, \quad (6)$$

where $\Omega_R, \Omega_L$ are the rectangular matrices and $\Lambda$ is diagonal one.

For CFD applications, the explicit form of propagator matrix $A$ demands the very high memory of dimension about $M \times M$. For the moderate grids (for example, about 100 nodes over a single spatial coordinate) $M \sim 4\times 10^4$ in 2D case and $M \sim 5\times 10^6$ in 3D case. The decomposition of $A$ via a product of reduced matrices needs to store only $2M \times N + N$ numbers. In the 2-D case for $M \sim 4\times 10^4$ and $N \sim 40$, the memory saving is about three orders of magnitude. It should be stressed that calculations of snapshots by the reduced Schmid operator approximation $A_r = \Omega_R \Lambda \Omega_L$ and by DMD are equivalent:

$$u_k = (\Omega_R \Lambda \Omega_L)(\Omega_R \Lambda \Omega_L)...\Omega_R \Lambda(\Omega_L u_1) = (\Omega_R \Lambda \Omega_L)^{k-1} u_1 = \Omega_R \Lambda^{k-1} \Omega_L u_1 = \sum_{i=1}^{N} a_i v_i \lambda_i^{k-1}. \quad (7)$$

Thus, the DMD is implicitly based on the assumption of the Schmid operator diagonalizability that, in strict sense, constrains the applicability domain. Nevertheless, according to [7] any real matrix is diagonalizable in the generic choice. So, accounting for numerical errors, we may consider the operator $A$ to be diagonalizable with the probability 1, may be, in unstable manner.

*4. Relationship of phase angles and frequencies*

DMD provides the set of conjugated Koopman eigenvalues $\lambda_j = a_j \pm ib_j$ and corresponding eigenvectors. The eigenvalues may be recast as

$$\lambda_j = \left|\lambda_j\right| e^{\pm i\alpha_j} = \left|\lambda_j\right|(\cos\alpha_j \pm i\sin\alpha_j) = \left|\lambda_j\right| e^{\pm i\omega_j\Delta t}. \quad (8)$$

Where $\left|\lambda_j\right| = e^{\sigma_j\Delta t}$, $\alpha_j$ is a phase angle and the values $\omega_j = \alpha_j / \Delta t$ are usually interpreted as the frequencies. However, the phase angle is defined in nonunique manner due to a periodicity ($\alpha_j + 2\pi m$). Moreover, the angles $(\alpha_j, -\alpha_j)$ and $(\alpha_j - 2\pi, 2\pi - \alpha_j)$ correspond to the same couple of eigenvalues and are obtained by the rotation in opposite directions. Thus, this phase angles engender quite different frequencies

$$\begin{aligned} &(\omega_j^{(1)}, -\omega_j^{(1)}) = (\alpha_j / \Delta t, -\alpha_j / \Delta t), \\ &(\omega_j^{(2)}, -\omega_j^{(2)}) = ((\alpha_j - 2\pi)/\Delta t, (2\pi - \alpha_j)/\Delta t). \end{aligned} \quad (9)$$

So, the single set of snapshots and, correspondingly, eigenvalues is not sufficient for the unique determination of the frequencies, that limit the DND applicability range.

Quite naturally, the estimation of $\omega_j$ may be performed via a comparison of two sets of snapshots obtained for close time intervals between snapshots ($\Delta t$ and $\Delta t + \delta t$, $\Delta t >> \delta t$), which formally provide adjacent phase angles $\alpha_k(\Delta t)$ and $\alpha_k(\Delta t + \delta t)$ from $\lambda_k(\Delta t)$ and $\lambda_k(\Delta t + \delta t)$.

Unfortunately, the direct numerical differentiation

$$\omega_k = \delta\alpha_k / \delta t. \quad (10)$$

was found to be extremely unstable in nonlinear event, mainly due to the variation of the number of complex eigenvectors even at small $\delta t$.

In an alternative approach, herein, we fixed the eigenvectors and eigenvalues determined from the first set of snapshots $u_k^{(1)}$. The comparison of the second set of snapshots $u_k^{(2)}$ and the forecast $\tilde{u}_k(\Delta t + \delta t) = \sum_j a_j v_j e^{\sigma_j (k-1)\Delta t} e^{i\omega_j (k-1)(\Delta t + \delta t)}$ enables to compare all combinations of $\omega_k$ using the mismatch functional

$$\varepsilon = \sum_k \sum_j (u_{k,j}^{(2)} - \sum_i a_i v_{i,j} e^{\sigma_j (k-1)\Delta t} e^{i\omega_j (k-1)(\Delta t + \delta t)}) \cdot (u_{k,j}^{(2)} - \sum_i a_i v_{i,j} e^{\sigma_j (k-1)\Delta t} e^{i\omega_j (k-1)(\Delta t + \delta t)})^*. \quad (11)$$

The couples of frequencies $(\omega_j^{(1)}, -\omega_j^{(1)})$ and $(\omega_j^{(2)}, -\omega_j^{(2)})$ were selected to minimize discrepancy (11) using the item-by-item examination. At this sorting, every combination of frequencies was assumed to correspond to some binary number of the length $N_{compl}$ bits ($N_{compl}$ is the number of complex eigenvalues). For all these numbers, the mismatch (11) was computed and the discrepancies for $\omega_j^{(1)}(\Delta t+\delta t)$, $-\omega_j^{(1)}(\Delta t+\delta t)$ and $\omega_j^{(2)}(\Delta t+\delta t)$, $-\omega_j^{(2)}(\Delta t+\delta t)$ were compared. The set of frequencies, providing minimum for $\varepsilon(\Delta t+\delta t)$, was obtained as a mixture of frequencies $\omega_j^{(1)}$ and $\omega_j^{(2)}$, corresponding the rotations of eigenvalues in opposite directions. The non-uniqueness over $\alpha_j+2\pi m$ was not controlled that is a drawback of this algorithm. Fortunately, high harmonics were not observed in the test problem. For moderate number of complex eigenvectors ($N_{compl}$ to be about 10), this algorithm was found to be operational. Unfortunately, the computational expenses increase as $2^{N_{compl}}$ that restricts its applicability. At calculations for large number of snapshots (about 40) significant part of eigenvalues with the smallest $|\lambda_j|$ was neglected.

Despite certain advances of above algorithm, the numerical differentiation $\omega_k=\delta\alpha_k/\delta t$ is the only observable perspective for the analysis of the frequencies with account of all sources of ambiguity. So, its stable realization is highly desirable.

## 5. *Jet impingement simulation*

The numerical tests were conducted for the Schmid operator computed from $N$ snapshots, obtained from the numerical solution of the two dimensional Euler equations.

The Schmid operator was reconstructed via the product of the rectangular matrices $M\times N$, where $M=36000$ and $N=10\div 50$. Some results are provided below.

The oscillating flow modes are known to occur at a normal impingement of supersonic underexpanded jet on the plate [8,9]. In this mode, the shock wave structure causes the peripheral pressure maximums, which may lead to an unsteady separation. This flow pattern exists within a rather narrow range of flow parameters (Mach number $M_a$, pressure ratio $n=p_0/p_a$, distance from nozzle exit by the surface $x/d_a$, specific heat ratio $\gamma=C_p/C_v$ ). The results of the oscillating flow computation seem to be appropriate as a toy problem. Surely, certain influence of turbulence is neglected in the present approach. However, the shock induced unsteady separation bubble may be successfully modeled by an inviscid numerical method providing a good agreement with the experimental data [9]. So, this model correctly represents a true nonlinear unsteady flow dynamics.

Herein, the results of computation by 2D+1 Euler equations are presented.

$$\frac{\partial\rho}{\partial t}+\frac{\partial(\rho U_k)}{\partial x^k}=0; \tag{12}$$

$$\frac{\partial(\rho U_i)}{\partial t}+\frac{\partial(\rho U_k U_i + P\delta_{ik})}{\partial x^k}=0; \qquad (13)$$

$$\frac{\partial(\rho E)}{\partial t}+\frac{\partial(\rho U_k h_0)}{\partial x^k}=0; \qquad (14)$$

Here $U_i$ are the velocity components, $h_0=(U_1^2+U_2^2)/2+h$, $h=\gamma e$, $e=RT/(\gamma-1)$, $E=\left(e+(U_1^2+U_2^2)/2\right)$ are enthalpies and energies (per unit volume), $P=\rho RT$ is the state equation.

The computations are performed in the spatial domain $\Omega=(0<x<X_{\max},0<y<Y_{\max})$ at time interval $(0<t<t_f)$ with the flow snapshot recorded at equally spaced time subintervals $\Delta t$.

At the left boundary $(x=0)$, we accept the supersonic inflow conditions, corresponding a nozzle exit section, and the environment conditions (pressure, temperature, zero normal derivatives of velocities) on another part of the boundary. At the right boundary no penetration condition is set. On the lateral boundaries $(y=0, y=Y_{\max})$ we impose the outflow conditions in the supersonic region and the environment conditions at the subsonic part of the boundary.

The Euler equations were solved by a method of second order spatial accuracy [10] with the numerical fluxes calculated via the method by Sun and Katayama [11] and a second order time discretization.

Figure 1 provides the surface pressure variation in time at the axis of symmetry.

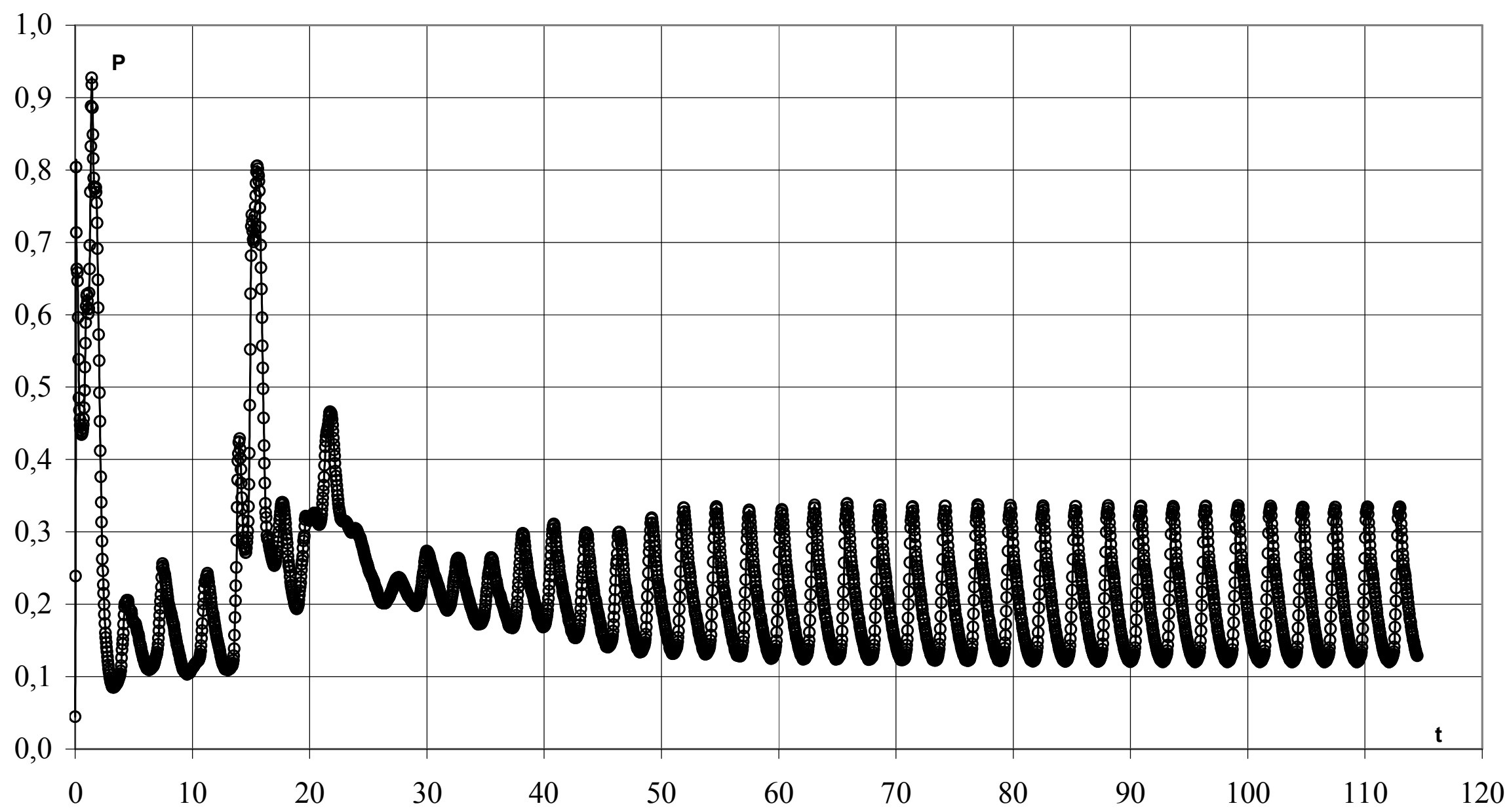

Figure 1. The pressure at the axis of symmetry as a function of dimensionless time.

Figure 2 presents details of the process in coordinates $(P,k)$, where $k$ is the number of CFD code steps. The periodic formation and disappearance of a separation bubble is specific for this mode. Figures 3 and 4 demonstrate the density fields for the maximal and minimal (developed separation bubble) pressure and corresponding streamlines. The results correspond to the flow parameters $M_a = 4.0$, $\gamma = 1.4$, $x/d_a = 15$, $n = 4$.

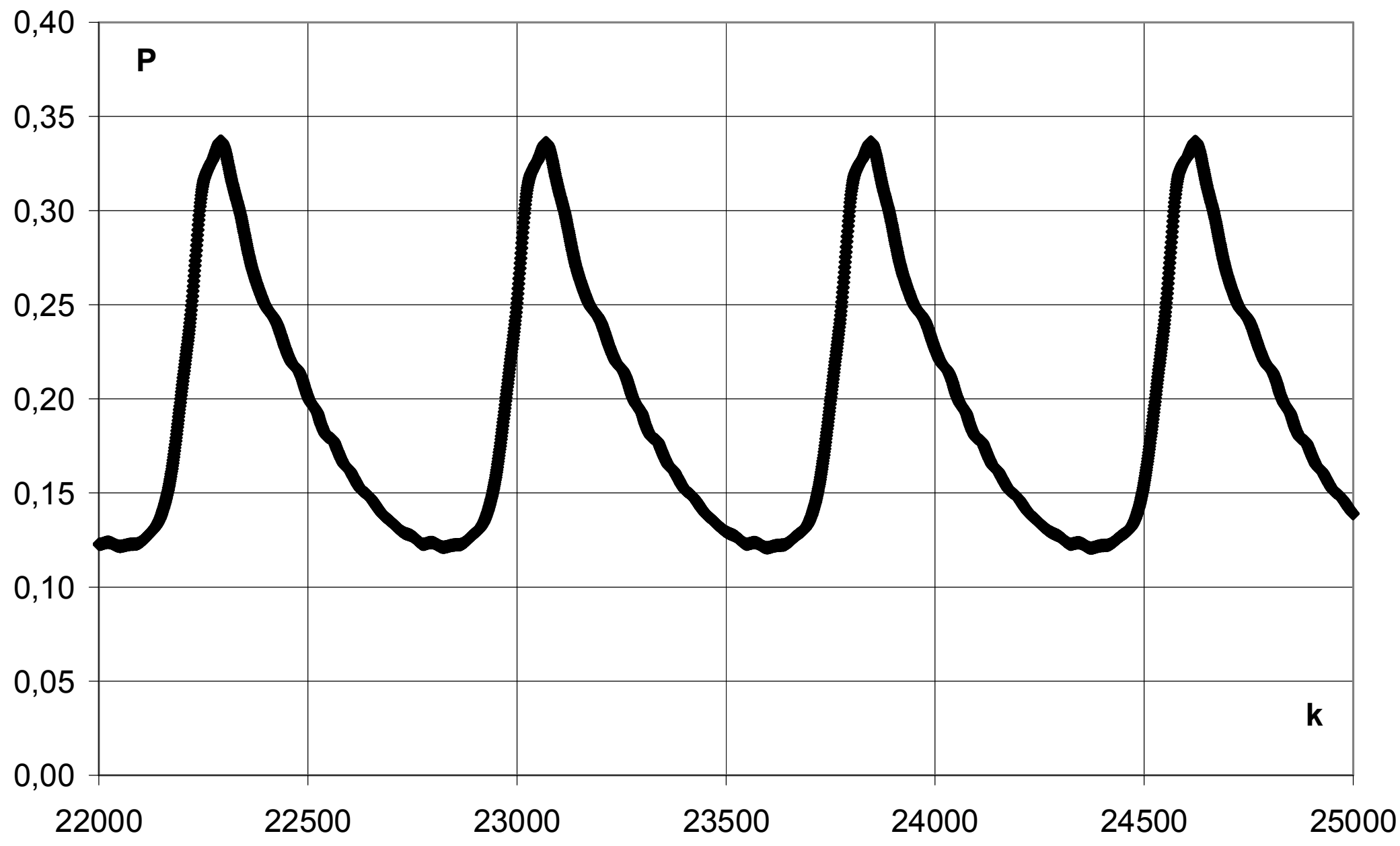

Figure 2. The pressure at the axis of symmetry as a function of number of computation steps.

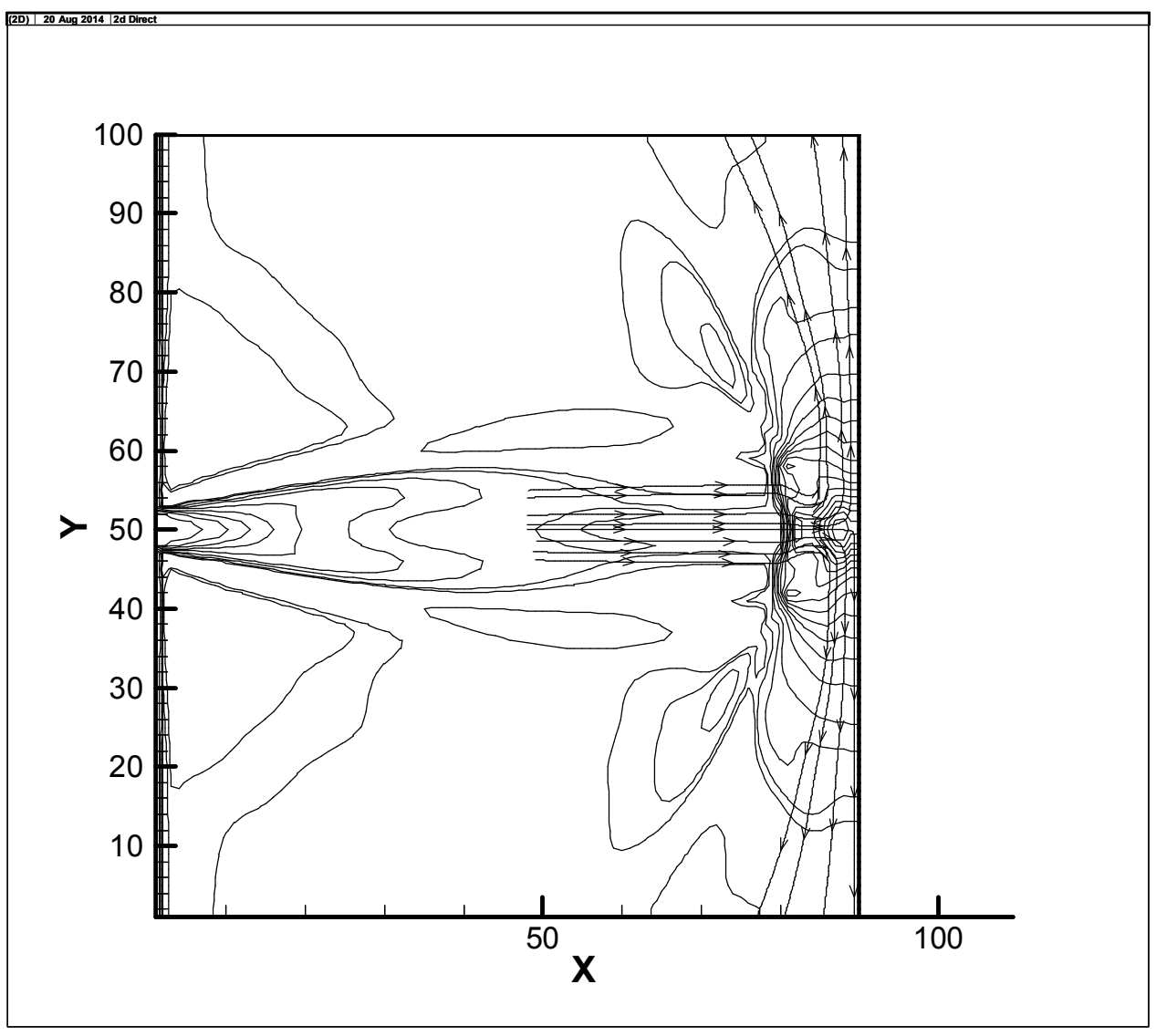

Figure 3. Density isolines and streamlines for the maximal pressure.

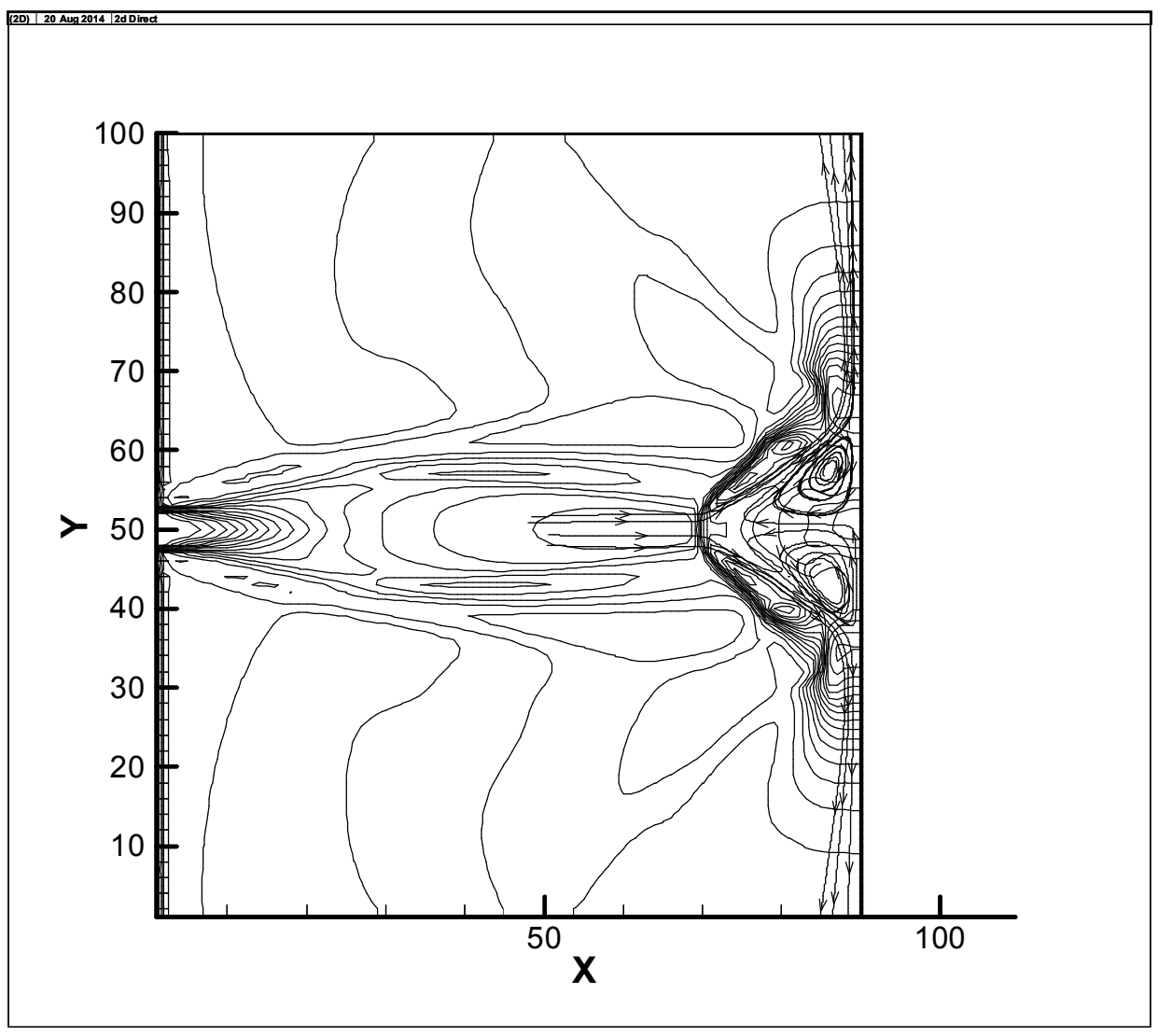

Figure 4. Density isolines and streamlines for the minimal pressure.

### 6. *DMD analysis of frequencies*

Herein, only self-oscillating part of the flow history is used for tests. Figure 5 presents the dependence $\lambda_{\mathrm{Re}}(\lambda_{\mathrm{Im}})$ for time interval $\Delta t = 500$ steps which far deviates from the oscillation period ($T_{osc} \approx 780$ steps). In this test 40 snapshots are used. Eigenvalues are located on the unit circle that is standard for nonlinear problems. The change of spectrum structure (number of complex eigenvalues) is found even for a small variation of $\Delta t$.

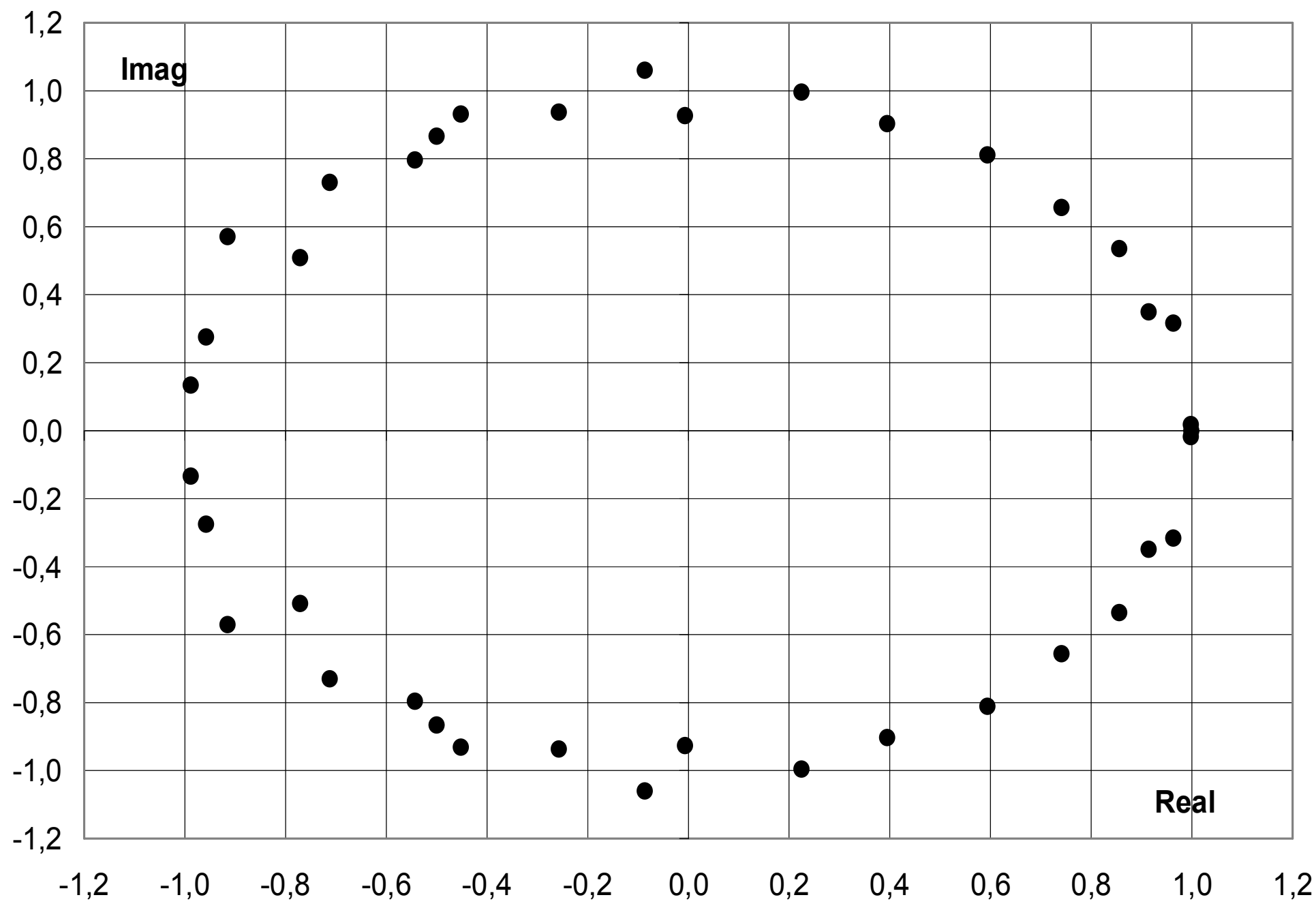

Figure 5. Eigenvalues $\lambda_{\mathrm{Re}}(\lambda_{\mathrm{Im}})$ for nonlinear mode.

The phase angle $\alpha_k(\lambda_k)$ does not provide the unique estimation for a frequency, as it was mentioned in Section 4. The amplitudes $|a_k(\omega_k)|$ are presented in Figure 6 in the dependence on minimal $\omega_j^{(1)}$ and maximal $\omega_j^{(2)}$ frequencies (9) for the data of Figure 5. One may see the messy spectrum even without account of the possible shift by $2m\pi$. The visually observed (Figures 1,2) main mode is not even included in the set of $\omega_j^{(1)}$. So, the difficulties in the identification of the relevant modes are evident.

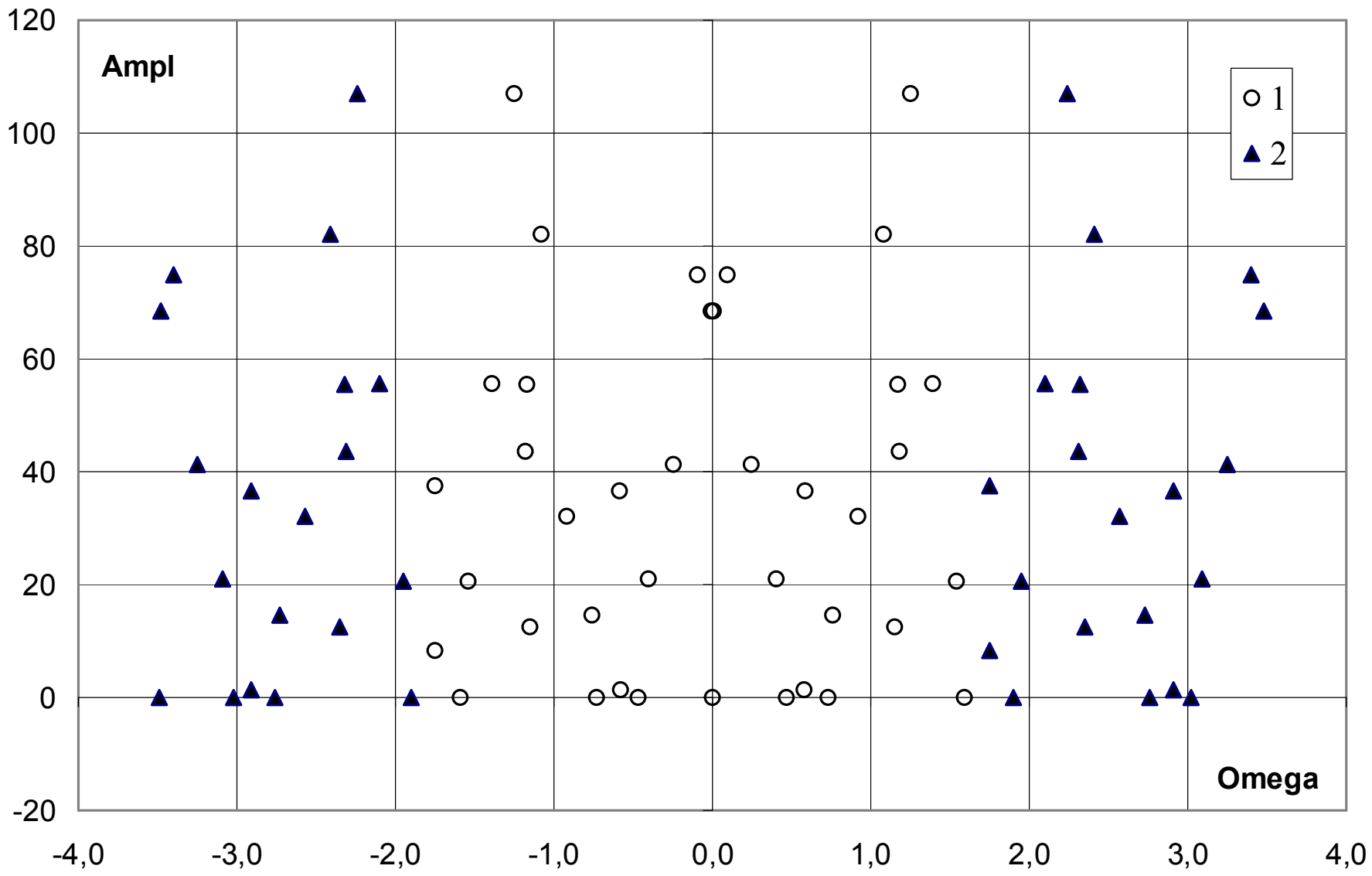

Figure 6. Amplitudes in dependence on the raw frequency estimations.

1- minimal frequency $\omega_j^{(1)}$, 2-maximum frequency $\omega_j^{(2)}$.

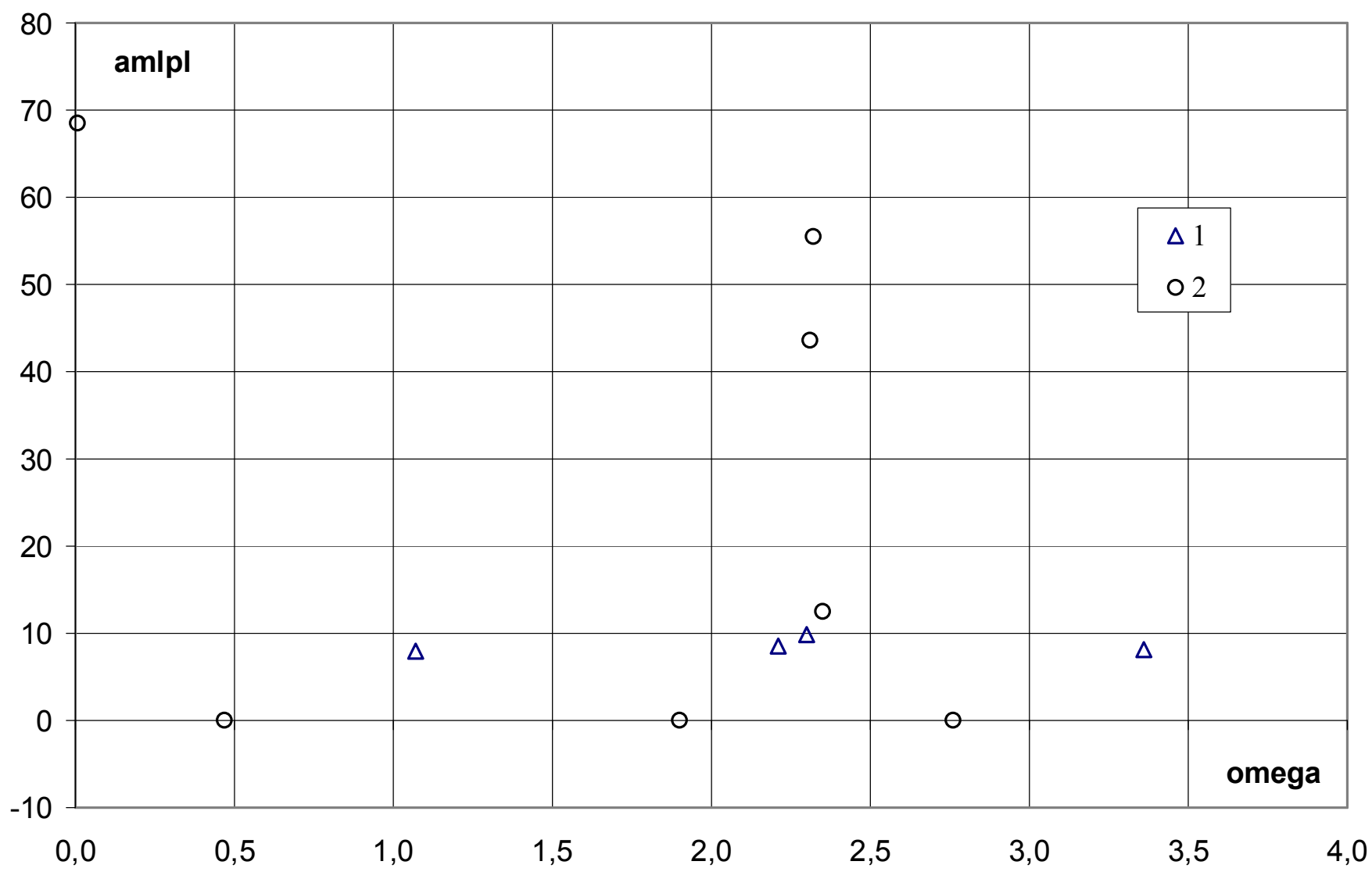

Figure 7. Amplitudes in dependence on the optimal frequency estimations (11).

1 corresponds 10 snapshots, 2 corresponds 40 snapshots.

Equation (11) enables to find the frequencies $\omega_j$ by comparing two data sets for the angles $\alpha_k(\Delta t)$ and $\alpha_k(\Delta t + \delta t)$ at $(\Delta t >> \delta t)$. Figure 7 shows the dependence of the amplitude on the frequency for 10 (line 1) and 40 (line 2) snapshots obtained by (11). Only positive frequencies are presented. The maximal amplitude correlates with the main frequency of oscillating mode ($\omega \approx 2.3$).

The results for 40 snapshots are obtained by saving 10 couples of the complex eigenvalues having maximum module and by neglecting the others. So, it may be seen, that the algorithm, based on Eq. (11), operates, at least for the moderate number of complex eigenvectors. Unfortunately, a computational cost increases as $2^{N_{compl}}$ that limits the applicability of this algorithm. Due to this circumstance (and due to unresolved problem of nonuniqueness at the shifts by $2\pi n$) a regularization of the problem that may provide stable eigenvalues and the consequent differentiation of phase angles, is highly desired.

## *7. Conclusion*

Koopman eigenvalues determine frequencies nonuniquelly due to the lack of account for a rotation direction and shifts by period.

The comparison of couple of snapshot sets corresponding close time intervals enables the unique estimation of frequencies.